\documentclass[12pt]{article}
\begin{document}
\begin{flushright}
CERN-TH/2001-341 \\
SLAC-PUB-9069 \\
hep-ph/0111390 \\
November 2001
\end{flushright}
\vspace*{1.cm}
\begin{center}
{ \large \bf High-Energy Asymptotics of \\  
   Photon--Photon Collisions in QCD}
\footnote{Presented by VTK at PHOTON2001, Ascona, Switzerland, 
2--7 September 2001, to appear in the Proceedings.} \\
\end{center}
\vspace*{0.3cm}
\begin{center}
{ \large Stanley~J.~Brodsky${}^{\$}$, Victor~S.~Fadin${}^{\dagger}$,
Victor~T.~Kim${}^{\ddagger \&}$, \\
 Lev~N.~Lipatov${}^{\ddagger}$
{\rm and}
Grigorii~B.~Pivovarov${}^{\S}$ } \\ 
\end{center}
\begin{center}
${}^\$ $  SLAC, Stanford, CA 94309, USA \\
${}^\dagger$  Budker Institute for Nuclear Physics,
Novosibirsk 630090, Russia \\
${}^\ddagger$ St.  Petersburg Nuclear Physics Institute,
Gatchina 188300, Russia \\
${}^\&$  CERN, CH-1211, Geneva 23, Switzerland \\
${}^\S$  Institute for Nuclear Research, Moscow 117312, Russia
\end{center}
\vspace{0.5cm}
\begin{center}
{\large \bf Abstract}
\end{center}
The high-energy behaviour of the total cross
section for highly virtual photons, as predicted by the
BFKL equation at next-to-leading order (NLO) in QCD, is presented.
The NLO BFKL predictions, improved
by the BLM optimal scale setting, are in excellent agreement
with recent OPAL and L3 data at CERN LEP2.

\newpage

Photon--photon collisions, particularly 
$\gamma^* \gamma^*$ processes, play a special role 
in QCD~\cite{Budnev75}, since their
analysis is much better under control than the calculation of hadronic
processes which require the input of non-perturbative hadronic
structure functions or wave functions.  In addition,
unitarization (screening) corrections due to multiple
Pomeron exchange should be less important
for the scattering of $\gamma^*$ of high
virtuality than for hadronic collisions.

The high-energy asymptotic behaviour of the $\gamma \gamma$ total cross
section in QED can be calculated~\cite{Gribov70} by an all-orders
resummation of the leading terms:
$\sigma \sim \alpha^4 s^{\omega}$, $\omega =
\frac{11}{32} \pi \alpha^2
\simeq 6 \times 10^{-5}$.  However, the slowly rising
asymptotic behaviour of the QED cross section is
not apparent since large contributions come
from other sources, such as the cut of the fermion-box
contribution:
$\sigma \sim \alpha^2 (\log s)/s$ \cite{Budnev75}
(which although subleading in energy dependence, dominates the rising
contributions by powers of the QED coupling constant),
and QCD-driven processes.

The high-energy asymptotic behaviour of hard QCD processes
is governed by the Balitsky--Fadin--Kuraev--Lipatov (BFKL) formalism
\cite{FKL,BL78}.
The highest eigenvalue, $\omega$, of the BFKL
equation \cite{FKL} is
related to the intercept of the QCD BFKL Pomeron,
which in turn governs the high-energy asymptotics
of the cross sections: $\sigma \sim
s^{\alpha_{I \negthinspace P}-1} = s^{\omega}$.
The BFKL Pomeron intercept in the leading order (LO)
turns out to be rather large:
$\alpha_{I \negthinspace P} - 1 =\omega_{LO} =
12 \, \ln2 \, ( \alpha_S/\pi )  \simeq 0.55 $ for
$\alpha_S=0.2$ \cite{FKL}.
The next-to-leading order (NLO)
corrections to the BFKL intercept have 
recently been calculated
\cite{FL}, but the results in the
$\overline{\mbox{MS}}$--scheme have a strong renormalization scale
dependence.
In Ref.\cite{BFKLP} we used the Brodsky--Lepage--Mackenzie (BLM)
optimal scale setting procedure~\cite{BLM} to eliminate the
renormalization scale ambiguity.  (For other approaches to the NLO BFKL
predictions, see Refs.\cite{Ciafaloni99,BFKLP} and references therein.)
The BLM optimal scale setting resums the conformal-violating
$\beta_0$-terms into the running coupling in all orders
of perturbation theory, thus preserving the conformal properties
of the theory.  The NLO BFKL predictions, as improved by 
the BLM scale setting, yields 
$\alpha_{I \negthinspace P} - 1 =\omega_{NLO} =$
0.13--0.18 \cite{BFKLP}.

The photon--photon cross sections with LO BFKL resummation
was considered in Refs.~\cite{BL78,Bartels96,Brodsky97}. 
Although the NLO impact factor of the virtual photon is not
known,  one can use the LO impact factor of 
\cite{Gribov70,BL78,Brodsky97},
assuming that the main energy-dependent 
NLO corrections come from the NLO
BFKL subprocess rather than the photon impact
factors \cite{KLP,BFKLP01}.

\begin{figure}[htb]
\begin{center}
\includegraphics{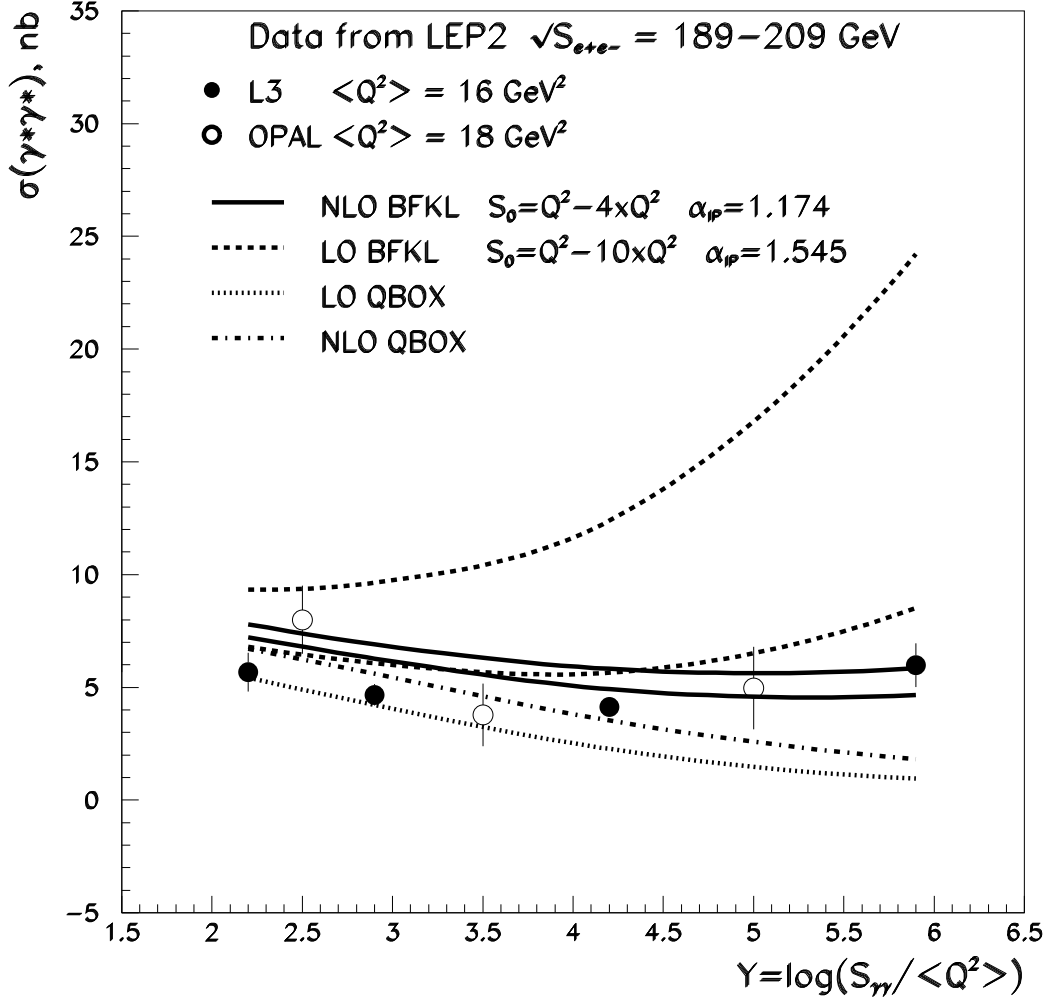}
\caption[*]{
The energy dependence of the total cross section for
highly virtual photon--photon collisions predicted
by the NLO BFKL theory\cite{KLP,BFKLP01,BFKLP} 
compared with OPAL\cite{OPAL} and L3\cite{L3} 
data from LEP2 at CERN.
The solid curves correspond to the BLM scale-fixed NLO BFKL predictions.
The dashed curve shows the LO BFKL prediction.  (Both predictions
include the quark-box contribution).  The BFKL predictions are shown
for two different choices of the Regge
scale, LO BFKL: $s_0= Q^2$--$10 Q^2$, NLO BFKL: $s_0=Q^2$--$4 Q^2$. }
\end{center}
\label{fig}
\end{figure}

Fig~\ref{fig} compares the LO and BLM scale-fixed NLO BFKL
predictions 
$\sigma \sim \alpha^2 \alpha_S^2 s^{\omega}$~\cite{BFKLP,KLP,BFKLP01}
with recent LEP2 data from OPAL\cite{OPAL}
and L3\cite{L3}.  The spread in the curves
reflect the uncertainty in the
choice of the Regge scale parameter, which defines
the beginning of the asymptotic regime:
 $s_0=Q^2  ~ {\rm to} ~ 10Q^2 $ for LO BFKL
and $s_0 = Q^2 ~{\rm to} ~ 4Q^2$ for NLO BFKL, where $Q^2$ is the mean
virtuality of the colliding photons.  One can see from Fig. \ref{fig}
that the agreement of the NLO BFKL predictions \cite{KLP,BFKLP01,BFKLP}
with the data is quite good.  We also note that the NLO BFKL
predictions are
consistent \cite{BFKLP01} with data recently presented
by ALEPH\cite{ALEPH}.
In contrast, the
NLO quark-box contribution \cite{Cacciari}
underestimates the L3 data point at 
$Y\equiv\log(s_{\gamma\gamma}/ \langle Q^2 \rangle)=6$
by more than 3 standard deviations.
The sensitivity of the NLO BFKL results
to the Regge parameter $s_0$ is much smaller than in
the case of the LO BFKL.  The variation of the predictions in the value of
$s_0$ reflects uncertainties from uncalculated
subleading terms.  The parametric variation of the LO BFKL predictions
is so large that it can neither be ruled out
nor confirmed at the energy range of LEP2.
\vspace*{1.0cm}

\vspace*{12.5cm}
The NLO BFKL phenomenology is consistent with the assumption of
small unitarization corrections in the photon--photon
scattering at large $Q^2$.  Thus one can accommodate the NLO BFKL
Pomeron intercept value 1.13--1.18 \cite{BFKLP} predicted by BLM
optimal scale setting.  In the case of hadron scattering, 
the larger unitarization corrections~\cite{Kaidalov86}
lead to a smaller effective Pomeron intercept value, 
about 1.1~\cite{Cudell00}.

In summary, highly virtual photon--photon collisions
provide a very unique opportunity to test high-energy asymptotics
of QCD.  The NLO BFKL predictions for the $\gamma^* \gamma^*$ 
total cross section, with the renormalization scale fixed
by the BLM procedure, show good agreement with the recent data from
OPAL\cite{OPAL} and L3\cite{L3} at CERN LEP2.

VTK thanks the Organizing Committee of PHOTON2001
for their kind hospitality and support.
This work was supported in part by the Russian
Foundation for Basic Research (RFBR), INTAS Foundation, and 
the U.S. Dept. of Energy
under contract No. DE-AC03-76SF00515.

\end{document}